\documentclass[a4paper,12pt]{article}

\usepackage{graphics,amsmath,amssymb}

\begin{document}

\title{
%\begin{flushright}
%{\normalsize IRB-TH-??/08}
%\end{flushright}
%\vspace{2 cm}
\bf Renormalization group scale-setting from the action - a road to modified gravity theories  
}

\author{Silvije Domazet\thanks{sdomazet@irb.hr} and Hrvoje \v Stefan\v ci\'c\thanks{shrvoje@thphys.irb.hr}}

\vspace{3 cm}
\date{
\centering
Theoretical Physics Division, Rudjer Bo\v{s}kovi\'{c} Institute, \\
   P.O.Box 180, HR-10002 Zagreb, Croatia}

\maketitle

\abstract{The renormalization group (RG) corrected gravitational action in Einstein-Hilbert and other truncations is considered. The running scale of 
the renormalization group is treated as a scalar field at the level of the action and determined in a scale-setting procedure recently introduced by 
Koch and Ramirez for the Einstein-Hilbert truncation. The scale-setting procedure is elaborated  for other truncations of the gravitational action and applied 
to several phenomenologically interesting cases. It is shown how the logarithmic dependence of the Newton's coupling on the RG scale leads to exponentially 
suppressed effective cosmological constant and how the scale-setting in particular RG corrected gravitational theories yields the effective $f(R)$ modified 
gravity theories with negative powers of the Ricci scalar $R$. The scale-setting at the level of the action at the non-gaussian fixed point in Einstein-Hilbert 
and more general truncations is shown to lead to universal effective action quadratic in Ricci tensor. 
%Recently obtained analytical solutions for the quadratic action in $R$ are summarized as an illustration of the dynamics at the non-gaussian fixed point.
}

%\vspace{2cm}

\section{Introduction}

% In the last fifteen years modern cosmology has reshaped our picture of the universe \cite{a17,a18,a19}. A very large amount of data from a stream of observations and 
% experiments \cite{Spergel:2006hy,Knop:2003iy,Riess:2004nr} has revealed that the nature of most of matter and/or interactions driving the dynamics of the universe is not understood. A particular 
% challenge is the source of the late-time phase of accelerated expansion that the universe is currently undergoing. Big theoretical efforts 
% \cite{Copeland:2006wr,Frieman:2008sn,Padmanabhan:2008if} have still 
% not lead to unequivocal theoretical descriptions of the cause for this accelerated expansion. Proposals concentrate on two fronts: the existence of novel component 
% with negative pressure, called dark energy, and modifications of gravitational interaction at cosmological scales.
% 
% Theories of modified gravitation have been attracting a lot of attention in recent years as a promising route towards elucidation of a mechanism producing the 
% accelerated expansion of the universe. A fast growing body of literature has been devoted to postulation of various modified gravity action functionals, 
% and the study of their cosmological and astrophysical implications \cite{modgrav,DeFelice:2010aj,Sotiriou:2008rp,Appleby,OdintsovRev,DeLaurentis,CliftonRev}. Numerous proposals of this kind were severely constrained by the available data. Yet, certain 
% modified gravity models still remain contenders in the quest for the explanation of the late-time cosmic acceleration.
% 
The discovery of the fascinating phenomenon of late-time accelerated cosmic expansion and detailed studies of effects attributed to dark matter have given a new impetus to 
investigation of quantum effects on curved background as a possible source of their explanation.   
%On the other hand, 
Quantum Field Theory (QFT) in curved space-time studies the connection of quantum phenomena and the background metric, allowing the investigation 
of influence of quantum fluctuations on the dynamics of space-time. Running (scale-dependent) parameters of the theory are one of principal results in QFT in curved 
space-time. At least two directions of research centered on the role of running parameters in the study of gravitational phenomena in cosmology and astrophysics 
have been developed in the last decade or so. The first is related to perturbative radiative corrections as a source of the scale dependence of the parameters of the 
theory \cite{Shapiro:1999zt,Shapiro:2000dz,Babic:2001vv,Babic:2004ev,ShapiroJCAP,Domazet:2010bk}.
%Guberina:2002wt,Gorbar:2002pw,Shapiro:2003ui,EspanaBonet:2003vk,Babic:2004ev,ShapiroJCAP,Bauer:2005rpa,EffDE,DynPar,LXCDM,DEpert,Shapiro:2009dh,Pert,Domazet:2010bk}. 
The second one is based on the concept of the average effective action accompanied by the machinery of the exact renormalization group 
\cite{Wetterich,Reuter1998,DouPercacci}.
%PercacciPerini,ReuterReview,WetterichReview,Bonanno:2001xi,Reuter:2003ca,Reuter:2005kb,Reuter:2004nv,Reuter:2004nx}. 
Some of the mentioned references also tried to develop methods useful for both approaches with scale dependent parameters \cite{Babic:2004ev,Domazet:2010bk}. 
In this paper we follow a similar philosophy.

A serious issue behind many, if not all, modified gravity theories is the fundamental origin of corrections to General Relativity (GR). 
%{\bf komentirati druge pristupe u kojima korekcije GR dolaze iz dobro fundiranih pristupa !!!}. 
Recently there have been several proposals how to relate the renormalization group (RG) effects to corrections to GR and modification of cosmological
dynamics in general \cite{HindmarshCosm,Contillo1,Contillo2,Contillo3,Bonanno:2012jy,Hindmarsh:2012rc}.
Our main goal in this paper is to systematically establish a connection between the QFT in curved space-time with its 
running (scale-dependent) parameters and modified gravity theories. 
% In particular, a procedure leading from the action of QFT in curved space-time with its 
% running parameters to equations 
% of motion with modified gravity will be laid out. We adopt a recently introduced approach of Koch and Ramirez \cite{Koch:2010nn}, generalize it and further 
% elaborate it through applications to new physical situations.
We base our approach on a procedure of scale-setting at the level of the action recently introduced by Koch and Ramirez \cite{Koch:2010nn}. Our contributions beyond 
\cite{Koch:2010nn} presented in this paper comprise a technical extension of the scale-setting procedure to the action with higher-derivative terms (section 2.2), 
a possible explanation of the smallness of the observed cosmological constant (section 3), a demonstration how the scale-setting procedure may result in $1/R^{\alpha}$ 
terms in the effective action (section 4) and the description of the universal form of the effective action at the non-gaussian fixed point in Einstein-Hilbert and 
other truncations (section 5).    

% The presentation of the paper is the following. After the introduction presented in the first section, the formalism of the scale-setting at the level of the action,
% both for Einstein-Hilbert truncation and action with higher order terms, is presented in the second section. In the third section the scale setting is demonstrated 
% on an example from QFT in curved space-time resulting in asymptotic de Sitter dynamics with an exponentially suppressed value of effective cosmological constant. 
% In the fourth section we present an effective energy-momentum tensor emerging from the scale-setting from the action and in the fifth section we present an 
% example of running of parameters in the Einstein-Hilbert truncation which results in the effective $f(R)$ action with negative powers of $R$. In the sixth section, 
% we focus on the scale-setting at the level of action at non-gaussian fixed points in Einstein-Hilbert and more general truncations. Finally, the seventh section 
% closes the paper with the discussion and the conclusions.           

\section{Scale-setting in quantum-corrected gravitational action}

The gravitational part of the action in general relativity with cosmological constant is described by the well-known Einstein-Hilbert action:
\begin{equation}
 \label{eq:GR}
S_{EH}=\int d^4 x \sqrt{-g}\frac{1}{16 \pi G} (R-2 \Lambda) \, ,
\end{equation}
with $\Lambda=8 \pi G \rho_{\Lambda}$. This action should be supplemented with the matter action $S_m$, i.e. the total action is $S=S_{EH}+S_m$.
Quantum corrections to this action, emanating from the fluctuations of quantum fields, modify the Einstein-Hilbert action 
in two principal ways: 1) additional higher-order terms in curvature must be added to remove the divergences in the process of renormalization and 2) the parameters of 
the terms in this enlarged action (such as $G$ and $\Lambda$) acquire scale dependence. 
The scale-dependent parameters can be introduced at the level of solutions, if they are available in analytical form, at the level of the EOM 
\cite{Babic:2004ev,ShapiroJCAP,Domazet:2010bk,PhiSett} or at the level of the action. In this paper we concentrate on the last of these approaches.

% The scale-dependent parameters can be introduced at three levels. In the cases where the analytical solutions of GR equations are available, these solutions can be 
% corrected by substituting parameters such as $G$ or $\Lambda$ with their scale-dependent counterparts. An analogous substitution can be performed at the level of 
% equations of motion. This approach was adopted in a number of papers, see e.g. \cite{Babic:2004ev,ShapiroJCAP,Domazet:2010bk,PhiSett}. 
% Finally, scale-dependent parameters can be introduced at the level of the action 
% itself. In this paper we concentrate on this approach.

Once the level where the scale-dependent (or running) parameters will be introduced is selected, an essential question is which value should be taken for the scale 
itself. As various arguments have been put forward in favor of different choices for the scale, a clear need for a systematic scale-setting procedure exists. 
For the approach based on correcting the equations of motion, the scale-setting procedure was proposed and applied in cosmology several years ago 
\cite{Babic:2004ev}, with a 
recent extension to space-times of lower symmetry with applications for astrophysical systems \cite{Domazet:2010bk}. 
The scale-setting procedure at the level of the action, 
recently proposed in \cite{Koch:2010nn}, opens ways for additional insights into the workings of quantum corrected gravitational theories.

\subsection{Einstein-Hilbert truncation}

In the Einstein-Hilbert truncation of the vacuum action of gravity, the parameters $G$ and $\Lambda$ become running parameters $G_k$ and $\Lambda_k$. i.e. they 
acquire scale dependence. In the approach of \cite{Koch:2010nn}, the running parameters are introduced at the level of the action:
\begin{equation}
 \label{eq:GRcorr}
S_{EH}=\int d^4 x \sqrt{-g}\frac{1}{16 \pi G_k} (R-2 \Lambda_k) \, .
\end{equation}

% The practical applicability of the action (\ref{eq:GRcorr}) requires the scale $k$ to be defined in terms of some other physical quantities. The importance of this 
% definition is unfortunately matched by the lack of clear physical argument for the identification of the correct choice for $k$. There have been many attempts to 
% assign physical importance to one choice or the other, but the argumentation in favor of each of such choices has so far, although in some cases well motivated, 
% remained qualitative. Fortunately, reasonable requirement that the choice of the scale $k$ should not violate symmetries of the theory yields a formal procedure 
% for determining its value. Namely, from the requirement of general covariance and Einstein tensor properties, a consistent procedure of scale-setting 
% at the level of the equation of motion has been defined and applied in cosmology and astrophysics \cite{Babic:2004ev,Domazet:2010bk}.

The identification of the scale obtained in the scale-setting procedure should be consistent with the original role of the scale, e.g. an infrared cutoff of an 
average effective action for the theory of gravity. The proof of this consistency might not be easily feasible. However, any other choice for scale other 
than the one obtained in 
the scale-setting procedure might be at odds with general covariance or require phenomenological extensions such as energy-momentum interchange with matter 
components. The scale-setting procedure represents a self-consistent way of determining the scale $k$.

The most general choice for the scale $k$ respecting the requirement of general covariance is for $k$ to become a scalar filed i.e. $k \rightarrow k(x)$, 
as proposed in \cite{Koch:2010nn}. This assumption has been recently incorporated into a number of lines of research \cite{Bonanno:2012jy,Hindmarsh:2012rc}
 in which an Ansatz for the identification of $k$ with some curvature invariants was used. A crucial advantage of the scale-setting procedure described below
is that, once the functional dependences of the action parameters on $k$ are known, the identification of the scale follows without additional assumptions. 
%This is a strong distinction compared to approaches where $k$ is chosen on grounds of reasonable, but qualitative arguments.

Following the presentation of \cite{Koch:2010nn}, the variation of the action (\ref{eq:GRcorr}) with respect to $k(x)$ yields the relation
\begin{equation}
\label{eq:setk}
R \left( \frac{1}{G_k} \right)'-2 \left( \frac{\Lambda_k}{G_k} \right)'=0 \, ,
\end{equation}
where primes denote differentiation with respect to $k$. The variation of the action (\ref{eq:GRcorr}) with respect to metric $g^{\mu \nu}$ yields an equation of 
motion 
\begin{equation}
\label{eq:gmunu} 
G_{\mu\nu} = -8 \pi G_k T_{\mu \nu} - \Lambda_k g_{\mu\nu} - \Delta t_{\mu\nu} \, ,
\end{equation}
where $G_{\mu \nu}=R_{\mu \nu} - 1/2 R g_{\mu \nu}$ is the Einstein tensor and $\Delta t_{\mu \nu} = G_k (\nabla_{\mu} \nabla_{\nu} - g_{\mu \nu} \Box) \frac{1}{G_k}$.
% \begin{equation}
%  \label{eq:deltat}
% \Delta t_{\mu \nu} = G_k (\nabla_{\mu} \nabla_{\nu} - g_{\mu \nu} \Box) \frac{1}{G_k} \, .
% \end{equation}
Taking the covariant derivative of (\ref{eq:gmunu}) and assuming the conservation of the energy-momentum tensor of matter $\nabla^{\nu} T_{\mu \nu}=0$, one obtains 
the relation
% \begin{equation}
%  \label{eq:covdev}
% G_{\mu \nu} \nabla^{\nu} \left( \frac{1}{G_k} \right) = - g_{\mu \nu} \nabla^{\nu} \left( \frac{\Lambda_k}{G_k} \right) - \nabla^{\nu} (\nabla_{\mu} \nabla_{\nu} -  
% g_{\mu \nu} \Box) \frac{1}{G_k} \, .
% \end{equation}
% Using the definition of Riemann tensor
% \begin{equation}
% \label{eq:Riemanndef}
% \nabla^{\nu} (\nabla_{\mu} \nabla_{\nu}  - g_{\mu \nu} \Box) \frac{1}{G_k} = (\nabla_{\nu} \nabla_{\mu}-\nabla_{\mu} \nabla_{\nu})\nabla^{\nu} 
% \frac{1}{G_k} = -R_{\mu \nu} \nabla^{\nu} \frac{1}{G_k} \, ,
% \end{equation}
% the relation (\ref{eq:covdev}) can be reformulated as
\begin{equation}
\label{eq:covdevreformulated} 
R \nabla_{\mu} \left( \frac{1}{G_k} \right) - 2 \nabla_{\mu} \left( \frac{\Lambda_k}{G_k} \right) = 0 \, , 
\end{equation}
which leads directly to (\ref{eq:setk}), using the scalar nature of $k$.

In this approach, from (\ref{eq:setk}) the curvature scalar $R$ can be expressed as a function of the scale $k$. The inversion of this function gives the
scale $k$ as a function of $R$. When the expression $k=k(R)$ is inserted into the gravitational field equation (\ref{eq:gmunu}), the obtained equations of motion 
resemble those falling to class of $f(R)$ theories\footnote{This is expected since the theory where $k$ is a scalar field is equivalent to Brans-Dicke theory with 
a potential and $\omega=0$} (see \cite{CliftonRev} and references therein). Alternatively, one can insert the result for $k(R)$ into the action (\ref{eq:GRcorr}) to obtain a 
$f(R)$ modified action.  
% The self-consistent determination of $k$ thus leads to a theory falling to a class of $f(R)$ theories. These modified gravity theories have recently attracted a lot 
% of attention owing to their appealing phenomenological properties \cite{modgrav,DeFelice:2010aj,Sotiriou:2008rp,Appleby,OdintsovRev,DeLaurentis,CliftonRev}. 
% Along with some theoretical problems and observational limitations, the physical origin 
% of additional terms in modified gravity actions remains the greatest challenge for $f(R)$ theories and modified gravity theories in general. 
The self-consistent 
determination of the scale $k$ through a scale-setting procedure represents a way to systematically introduce the modifications of gravity to equations of motion 
(\ref{eq:gmunu}) or the action (\ref{eq:GRcorr}).

\subsection{The action with higher-order terms}

For the reasons of renormalizability, the Einstein-Hilbert action in general needs to be completed with the higher-derivative terms to obtain the vacuum action of 
QFT in curved space-time \cite{Shapiro:2008sf}. Making a step beyond \cite{Koch:2010nn}, these additional contributions to the action
\begin{equation}
 \label{eq:HD}
S_{HD}=\int d^4 x \sqrt{-g} (a_1 C^2 + a_2 E + a_3 \Box R + a_4 R^2) 
\end{equation}
comprise the terms with the square of the Weyl term $C^2=R^{\mu \nu \lambda \tau} R_{\mu \nu \lambda \tau} - 2 R^{\mu \nu} R_{\mu \nu} + (1/3) R^2$, Gauss-Bonnet 
term with the integrand $E=R^{\mu \nu \lambda \tau} R_{\mu \nu \lambda \tau} - 4 R^{\mu \nu} R_{\mu \nu} + R^2$ and terms with $\Box R$ and $R^2$. 
An interesting study of the issue of ghosts in the action (\ref{eq:HD}) was recently given in \cite{Mazumdar}. The inclusion of 
quantum corrections arising in QFT makes the coefficients scale-dependent i.e. $a_i \rightarrow a_{i,k}$ for $i=1,2,3,4$. 
% Then the complete vacuum action takes the 
% form
% \begin{eqnarray}
% \label{eq:EHHD}
% S&=&S_{EH}+S_{HD} \nonumber \\
% &=&\int d^4 x \sqrt{-g}\left[ \frac{1}{16 \pi G_k} (R-2 \Lambda_k)+a_{1,k} C^2 + a_{2,k} E + a_{3,k} \square R + 
% a_{4,k} R^2 \right] \, .
% \end{eqnarray}
%
Using the same argumentation as above, we can assume the scale $k$ to be a scalar field and the variation of the total action $S=S_{EH}+S_{HD}$
%(\ref{eq:EHHD}) 
with respect to the scale $k$ then becomes
\begin{equation}
 \label{eq:varEHHDk}
\frac{1}{16 \pi}\left( \left(\frac{1}{G_k}\right)' R-2 \left(\frac{\Lambda_k}{G_k}\right)' \right)+a_{1,k}' C^2 + a_{2,k}' E + a_{3,k}' 
\square R + a_{4,k}' R^2 = 0 \, ,
\end{equation}
where $'$ denotes differentiation with respect to $k$. This equation formally connects the scale $k$ with the invariants $R$, $R^2$, $\square R$, $C^2$ and $E$. 
Solving for $k$ in terms of the said invariants %from (\ref{eq:varEHHDk}) 
and introducing this $k$-dependence into the gravitational equations of motion obtained 
by variation of $S=S_{EH}+S_{HD}$ with respect to $g_{\mu \nu}$, one obtains the equations of motion that resemble those of modified gravity theory with the terms in 
action dependent on $R$, $R^2$, $\square R$, $C^2$ and $E$.
%{\bf although HD terms might not be important in EOS they might be important in scale-setting}
It is interesting to notice that the variation of $E$ term in (\ref{eq:HD}) does not reduce to total derivative owing to space-time dependence of $k$, 
and consequently of $a_{2,k}$.

% Finally, the coupling constants in the part of the action describing matter, $S_{matt}$ should also acquire scale dependence. For the full 
% action $S=S_{EH}+S_{HD}+S_{matt}$, the variation of the action with respect to $k$ yields a scale-setting relation implying that $k$ should also depend on matter field 
% invariants along with metric field invariants.  

%The scale-setting procedure described so far will be further on applied to some physically interesting examples and its general consequences will be discussed. 

\section{Scale-setting for a small effective cosmological constant}

In the considerations of the running of $\rho_{\Lambda}$ and $G$ from QFT in curved space-time \cite{ShapiroJCAP}, 
the running of the cosmological constant energy density has the quadratic dependence on the scale
\begin{equation}
 \label{eq:qftrholam}
\rho_{\Lambda,k}=c_0+c_2 k^2 \, ,
\end{equation}
where $c_0$ and $c_2$ are real constants, whereas the Newton's coupling runs logarithmically with the scale
\begin{equation}
 \label{eq:qftG}
G_k=\frac{G_0}{1+d_2 \ln \frac{k^2}{k_0^2}} \, .
\end{equation}
Here $d_2$ and $G_0$ are real constants.
An insertion of (\ref{eq:qftrholam}) and (\ref{eq:qftG}) into the scale-setting relation (\ref{eq:setk}) readily yields a simple expression for the scale in terms of the 
Ricci scalar
\begin{equation}
\label{eq:qftscale}
k^2=\frac{d_2}{16 \pi c_2 G_0} R \, .
\end{equation}
With this value of the scale, the expressions for the CC energy density and the running Newton's coupling become
\begin{equation}
 \label{eq:qftrholam_R}
\rho_{\Lambda,k}=c_0+c_2 \chi R\, , \;\;\; G_k=\frac{G_0}{1+d_2 \ln \frac{R}{R_0}} \, ,
\end{equation}
% and
% \begin{equation}
%  \label{eq:qftG_R}
% G_k=\frac{G_0}{1+d_2 \ln \frac{R}{R_0}} \, ,
% \end{equation}
where $\chi=d_2/(16 \pi c_2 G_0)$. It is intriguing that the expression  for $G$  in (\ref{eq:qftrholam_R}) has been recently proposed on the basis of analogy 
with $\alpha_{QCD}$ running coupling constant in \cite{Frolov:2011ys}. Here it is derived starting from general arguments from QFT 
on curved space-time using the scale-setting procedure.

An interesting aspect of relations (\ref{eq:qftrholam_R}),
% and (\ref{eq:qftG_R}), 
i.e. their dependence on $R$, is the de Sitter regime that they allow in the 
maximally symmetric space-time. Namely, in the FLRW space-time, the $00$ component of the equation (\ref{eq:gmunu}) takes the form
\begin{equation}
 \label{eq:friedmann}
\left( \frac{\dot{a}}{a} \right)^2 = \frac{8 \pi G_k}{3} \left(\rho_{r,0} \left(\frac{a}{a_0}\right)^{-4} + \rho_{m,0} \left(\frac{a}{a_0}\right)^{-3} \right) 
+ \frac{\Lambda_k}{3} - \frac{\kappa}{a^2} + \frac{\dot{G_k}}{G_k} \frac{\dot{a}}{a} \, .
\end{equation}
Asymptotically when $a \rightarrow \infty$, using $R=12 H^2 + 6 \dot{H}$ in the spatially flat FLRW metric, the above equation acquires the form
%$H^2=\frac{\Lambda_k}{3}+H \frac{\dot{G_k}}{G_k}$
\begin{equation}
 \label{eq:asym}
H^2=\frac{\Lambda_k}{3}+H \frac{\dot{G_k}}{G_k} \, .
\end{equation}
To have a de Sitter regime, $H^2=\mathrm{const}$, $R=12 H^2=\mathrm{const}$, (\ref{eq:asym}) must take the form $H^2=\frac{8 \pi}{3} G_k \rho_{\Lambda,k}$.
% \begin{equation}
%  \label{eq:dS}
% H^2=\frac{8 \pi}{3} G_k \rho_{\Lambda,k} \, .
% \end{equation}
Inserting the expressions for $\rho_{\Lambda,k}$ and $G_k$ one obtains
\begin{equation}
 \label{eq:dS2}
H^2=\frac{\frac{8 \pi G_0}{3} c_0 +2 d_2 H^2}{1 + d_2 \ln \frac{12 H^2}{R_0}} \, .
\end{equation}
If $c_0$ is negligible and $H^2 \not= 0$, 
%(\ref{eq:dS2}) becomes
% \begin{equation}
% \label{eq:dS3}
% H^2 \left( 1- \frac{2 d_2}{1 + d_2 \ln \frac{12 H^2}{R_0}} \right)=0 \, .
% \end{equation}
% For $H^2 \not= 0$, 
(\ref{eq:dS2}) gives
\begin{equation}
 \label{eq:dS4}
H^2=\frac{R_0}{12} e^{\frac{2 d_2-1}{d_2}} \, .
\end{equation}
For $d_2 \ll 1$, the de Sitter scale $H$ is exponentially suppressed compared to the scale $R_0$, as already identified in \cite{Frolov:2011ys}.

\section{Power corrections in $R$}

A very interesting question is if the running laws for $\rho_{\Lambda,k}$ or $G_k$ in terms of the scale $k$ can be found which, by identifying $k$ in terms of 
$R$ through a scale-setting relation, in the Einstein-Hilbert truncation lead to power law corrections in $R$ \cite{Turner}. In this section we show that such running laws acquire relatively 
simple forms. For the running laws of the form
\begin{equation}
\label{eq:run1lam}
\rho_{\Lambda,k}=A_1+B_1 k^{\gamma} \, , \;\;\; \frac{1}{G_k}=A_2+B_2 k^{\delta} \, ,
\end{equation}
% and 
% \begin{equation}
% \label{eq:run1G}
% \frac{1}{G_k}=A_2+B_2 k^{\delta} \, , 
% \end{equation}
the scale-setting relation (\ref{eq:setk}) yields
\begin{equation}
 \label{eq:run1k}
k=\left( \frac{1}{16 \pi} \frac{\delta}{\gamma} \frac{B_2}{B_1} R  \right)^{\frac{1}{\gamma-\delta}} \, .
\end{equation}
Inserting this result for the scale $k$ into (\ref{eq:run1lam}) gives
% and (\ref{eq:run1G}) gives
\begin{equation}
\label{eq:run2rho}
\rho_{\Lambda,k}=A_1+B_1 \left( \frac{1}{16 \pi} \frac{\delta}{\gamma} \frac{B_2}{B_1} \right)^{\frac{1}{1-\delta/\gamma}} R^{\frac{1}{1-\delta/\gamma}} =
A_1+B_1 \left( \frac{1}{16 \pi} \frac{\alpha+1}{\alpha} \frac{B_2}{B_1} \right)^{-\alpha} \frac{1}{R^{\alpha}} \, ,
\end{equation}
% and 
\begin{equation}
\label{eq:run2G}
\frac{1}{G_k}=A_2+B_2 \left( \frac{1}{16 \pi} \frac{\delta}{\gamma} \frac{B_2}{B_1} \right)^{\frac{\delta/\gamma}{1-\delta/\gamma}} 
R^{\frac{\delta/\gamma}{1-\delta/\gamma}} = A_2+ 16 \pi \frac{\alpha}{\alpha+1} B_1 \left( \frac{1}{16 \pi} \frac{\alpha+1}{\alpha} 
\frac{B_2}{B_1} \right)^{-\alpha} \frac{1}{R^{\alpha+1}}\, , 
\end{equation}
where we used $\frac{1}{1-\delta/\gamma}=-\alpha \Rightarrow \frac{\delta}{\gamma}= \frac{\alpha+1}{\alpha}$.
%From (\ref{eq:run2rho}) and (\ref{eq:run2G}) it is immediately clear that they reproduce the power-law structure of (\ref{eq:rholamR}) and (\ref{eq:GkR}) for 
% \begin{equation}
%  \label{eq:alphadef}
% \frac{1}{1-\delta/\gamma}=-\alpha \Rightarrow \frac{\delta}{\gamma}= \frac{\alpha+1}{\alpha}\, .
% \end{equation}
For $\alpha > 0$ this is possible if $\delta/\gamma > 1$. There are two ways how this requirement on the ratio $\delta/\gamma$ can be satisfied: 
$\delta > \gamma > 0$ and $\delta < \gamma < 0$. It is important to notice that the negative powers of $R$ in the effective modified
gravity action can be obtained for positive powers of $k$ in (\ref{eq:run1lam}),
% and (\ref{eq:run1G}), 
i.e. for positive $\gamma$ and $\delta$. 
% From (\ref{eq:alphadef}) follows
% \begin{equation}
% \label{eq:defdelgam}
% \frac{\delta}{\gamma}= \frac{\alpha+1}{\alpha} \, .
% \end{equation}
% Using this definition, the relations (\ref{eq:run2rho}) and (\ref{eq:run2G}) can be written as 
% \begin{equation}
% \label{eq:run3rho}
% \rho_{\Lambda,k}=A_1+B_1 \left( \frac{1}{16 \pi} \frac{\alpha+1}{\alpha} \frac{B_2}{B_1} \right)^{-\alpha} \frac{1}{R^{\alpha}} 
% \end{equation}
% and 
% \begin{equation}
% \label{eq:run3G}
% \frac{1}{G_k}=A_2+ 16 \pi \frac{\alpha}{\alpha+1} B_1 \left( \frac{1}{16 \pi} \frac{\alpha+1}{\alpha} \frac{B_2}{B_1} \right)^{-\alpha} \frac{1}{R^{\alpha+1}} \, . 
% \end{equation}
With the identification $A_1=\rho_{\Lambda}^{*}$, $C=B_1 \left( \frac{1}{16 \pi} \frac{\alpha+1}{\alpha} \frac{B_2}{B_1} \right)^{-\alpha}$ and $A_2=1/G_{*}$, 
% the relations (\ref{eq:rholamR}) and (\ref{eq:GkR}) are fully reproduced.
the action containing power law terms in $R$ is obtained: 
\begin{equation}
\label{eq:actionR}
\frac{R-2\Lambda_k}{16 \pi G_k} = \frac{R}{16 \pi G_{*}} - \frac{1}{\alpha+1} \frac{C}{R^{\alpha}} - \rho_{\Lambda}^{*} \, .
\end{equation}
It is interesting to observe that for $A_1=0$ and $A_2=0$, the effective action of the form $\sim R^{-\alpha}$ is obtained.

\section{Scale setting at fixed points - universality in effective modified gravity theories}

In the asymptotic safety program \cite{Weinberg} (see also \cite{Percacci} for a recent review), an alternative to renormalizability of QFT theories, 
the existence of non-gaussian fixed point plays a key role. 
At non-gaussian fixed points (NGFP) the scaling of parameters in the actions, such as $G_k$ and $\Lambda_k$, with the cutoff $k$ is determined entirely from 
dimensionality of these parameters. Next we present our results that the scale-setting procedure results in modified gravity theories 
with some universal properties. 

\subsection{Non-gaussian fixed point in Einstein-Hilbert truncation}

In Einstein-Hilbert truncation (\ref{eq:GRcorr}) 
% \begin{equation}
% \label{eq:EH0}
% S_{\mathrm{EH}}=\int d^4 x \sqrt{-g} \frac{R-2 \Lambda_k}{16 \pi G_k} \, ,
% \end{equation}
the scaling of the Newton coupling and the cosmological constant at the non-gaussian fixed point are
\begin{equation}
\label{eq:EH1}
G_k=\frac{g^{*}}{k^2}\, , \;\;\;\; \Lambda_k=\lambda^{*} k^2 \, .
\end{equation}
The scale-setting condition (\ref{eq:setk}) then yields 
% \begin{equation}
% \label{eq:EH2}
% \frac{2 k}{g^{*}} (\partial_{\mu} k) (R-4 \lambda^{*} k^2)=0 \, .
% \end{equation}
% For $\partial_{\mu} k \not = 0$, follows
\begin{equation}
\label{eq:EH3}
k^2=\frac{R}{4 \lambda^{*}} \, .
\end{equation}
This leads to $\Lambda_k=\frac{R}{4}$ and $G_k=\frac{4 g^{*} \lambda^{*}}{R}$. Inserting these results into (\ref{eq:GRcorr}) yields a modified gravity action
\begin{equation}
\label{eq:EH4}
S_{\mathrm{EH}}=\int d^4 x \sqrt{-g} \frac{R^2}{128 \pi g^{*} \lambda^{*}} \, .
\end{equation}
Universality features of this action become clear in the absence of matter or in the regime where the influence of matter is negligible. 
In that case the exact values of constants $g^{*}$ and $\lambda^{*}$ are not important since they just enter a constant 
($\frac{1}{128 \pi g^{*} \lambda^{*}}$) multiplying the entire action and therefore do not affect the equations of motion. 
The behavior of the system near the non-gaussian fixed point for any $g^{*}$ and $\lambda^{*}$is then described by a $f(R)$ modified gravity 
theory with $f(R) = R^2$. The importance of $R^2$ effective action, which can be analytically solved \cite{Clifton:2007ih,Domazet:2012pt,Barrow}, was recently stressed in \cite{Bonanno:2012jy,Hindmarsh:2012rc}, based on Ansatz $ k^2 \sim R$.
In this paper the relation $k^2 \sim R$ is derived from the scale-setting procedure, together with the coefficient of proportionality.

\subsection{Non-gaussian fixed point in more general truncations}
For a more general action of the form
\begin{equation}
\label{eq:gen1}
S=\int d^4 x \sqrt{-g} \sum_{m=0}^n c_{k,m} R^m \, ,
\end{equation}
the scaling of the coefficients $c_{k,m}$ at a non-gaussian fixed point is $c_{k,m} = a_m k^{4-2 m}$.
% \begin{equation}
% \label{eq:gen2}
% c_{k,m} = a_m k^{4-2 m} \, .
% \end{equation}
The scale-setting condition (\ref{eq:setk}) then yields
\begin{equation}
\label{eq:gen3}
\sum_{m=0}^n (4-2 m) a_m \left( \frac{R}{k^2}\right)^m = 0 \, .
\end{equation}
If we define the polynomial $P(x)=\sum_{m=0}^n (4-2 m) a_m x^m$ and denote its zeros by $x_l$, $P(x_l)=0\, , (l=1,2,...,n)$, the result of the scale-setting 
procedure is
\begin{equation}
\label{eq:gen4}
k^2=\frac{R}{x_l} \, .
\end{equation} 

Inserting the result for $k$ into the action (\ref{eq:gen1}) gives
\begin{equation}
\label{eq:gen5}
S=\int d^4 x \sqrt{-g} R^2 \sum_{m=0}^n a_m x_l^{m-2} \, .
\end{equation}
The resulting action is the $R^2$ action. In the regime where the contribution of matter is negligible, the action exhibits universal properties at the 
non-gaussian fixed point. Namely, the particular values of the coefficients $a_m$ are immaterial since they all combine into a constant multiplying the entire 
action ($\sum_{m=0}^n a_m x_l^{m-2}$). Another interesting property is that this result is valid for any $n$, including the case when it becomes arbitrarily 
large i.e. when the polynomial in the action effectively becomes a series expansion in $R$.  

% \subsection{General power-law scaling} 
% 
% For the general power law scaling of the parameters in the Einstein-Hilbert action
% \begin{equation}
% \label{eq:genpow1}
% G_k=\frac{A_1}{k^{\beta}} \, , \;\;\;\; \Lambda_k=A_2 k^{\alpha} \, ,
% \end{equation}
% the scale-setting (\ref{eq:setk}) gives
% \begin{equation}
% \label{eq:genpow2}
% k=\left( \frac{\beta}{2(\alpha+\beta)} \frac{R}{A_2} \right)^{1/\alpha} \, .
% \end{equation}
% Putting this result back into (\ref{eq:genpow1}) results in the action
% \begin{equation}
% \label{eq:genpow3}
% S=\int d^4 x \sqrt{-g} \frac{1}{16 \pi A_1} \left( \frac{\beta}{2(\alpha+\beta) A_2} \right)^{\beta/\alpha} \frac{\alpha}{\alpha+\beta} R^{(\alpha+\beta)/\alpha}\, .
% \end{equation}
% 
% The effective action (\ref{eq:genpow3}) is of the $R^n$ type which has been extensively studied and analytical solutions of its corresponding equations of 
% motion have been obtained in \cite{Clifton:2007ih} and recently in \cite{Domazet:2012pt}.  
% %{\bf discussion on $R^{\alpha}$)}

\section{Discussion and conclusions}

Gravitational theories with renormalization group corrections all share the problem of determination of the RG scale. Results from the recent literature 
\cite{Bonanno:2012jy,Hindmarsh:2012rc,Frolov:2011ys} 
demonstrate the common notion that the RG scale should be related to some curvature invariant and the Ricci scalar seems to be the first choice. In these approaches, 
however, the scale is set ad hoc, frequently motivated by qualitative physical arguments. The scale-setting procedure at the level of action 
introduced in \cite{Koch:2010nn} and 
further elaborated and extended in this paper resolves the scale-setting problem by simple variation of the action over the RG scale. The dynamics of the theory then depends only 
on the running laws of the parameters of the action and no additional assumptions on the RG scale are needed.

Although the scale-setting procedure reproduces some ad hoc choices for the RG scale proposed in the literature, an important distinction of our results 
compared to these approaches is that in the scale setting procedure used in this paper the refinement of the running laws (e.g. by the addition of the higher-order 
terms in the $\beta$ functions) in general results in different values for the RG scale. In particular, although the scale-setting procedure at the NGFP 
yields $k^2 \sim R$ as proposed in \cite{Bonanno:2012jy,Hindmarsh:2012rc}, close to the NGFP the value of the scale deviates from this relationship as 
higher-order corrections are added. 
%This difference will hopefully provide sufficient basis for falsifiability compared to the mentioned approaches.

%The scale-setting procedure further brings important information on the behavior close to non-gaussian fixed points. 
In Einstein-Hilbert truncation, but also in 
truncations which are polynomials in $R$ of arbitrary powers, at NGFP the effective action obtained by the scale-setting procedure is described by the action quadratic in $R$. 
%It is intriguing that this universal $R^2$ form of effective action is obtained for actions which are polynomials in $R$ of arbitrary degree. 
Furthermore, if at NGFP it 
were possible to neglect the contributions of matter, the precise values of parameters would become irrelevant. Namely, they all combine into a constant multiplying
the entire action and therefore do not affect the dynamics of the system. In that case the universality of the functional form of the effective action is elevated
to the universality of the dynamics. 
%The solution of such dynamics in FLRW space-times is presented in section 6.1.

The research presented in this paper is concentrated on the gravitational sector of QFT in the curved space-time. There is no obstacle to extend the 
scale-setting procedure to matter sector since the principle behind the scale-setting, i.e. the variation of the action over the scale $k$ can be easily applied
to couplings and other parameters of the matter Lagrangian. Recent research \cite{Eichhorn} even suggests that such extension might be an important future step.

%\vspace{2cm}

{\bf Acknowledgements.}  
This work was
supported by the Ministry of Education, Science and Sports of the Republic of Croatia 
under the contract No. 098-0982930-2864.

\end{document}